# Impact of 700keV Ni++ ion irradiation on structural and optical properties of GaN

Aqeel Ahmed, Zaheer Ahmad

**Abstract:** In this paper, we present the effects of high-energy Ni++ ion irradiation on the structural and optical properties of GaN films. Three different irradiation doses of $10^{13}$, $10^{14}$, and $10^{15}$ ions/cm$^2$ were used while keeping the ion energy at 700keV. The irradiation induced structural and optical changes in GaN films were measured using X-ray diffraction (XRD) and UV-Vis spectroscopy. The XRD measurements on irradiated films discovered several extra peaks in the XRD spectrum compared with the as-grown GaN film indicating the formation of new phases or defects in the GaN film due to the ion irradiation. The intensity of these extra peaks increases with increasing ion dose, suggesting that the density of defects in the GaN film also increases with increasing ion dosage. The UV-Vis measurements revealed a decrease in the bandgap of the irradiated GaN films from 3.40 eV for the pristine GaN film to 3.26 eV for the film irradiated with highest dose of $10^{15}$ ions/cm$^2$. The decrease in bandgap can be attributed to the creation of defects and/or the formation of new phases in the GaN film due to the ion irradiation.

III-nitrides (GaN, InN and AlN) have attracted tremendous attention in recent decades and have been extensively studied[1,2,3,4,5,6] for their applications in light-emitting diodes (LEDs), solar cells, and power electronics. The performance of GaN-based devices is often limited by defects and impurities in the material. Therefore, it is imperative to investigate new ways to amend the optical, structural and transport properties of GaN films to improve their device performance. Ion irradiation is a useful tool for tailoring the crystalline strain that provides a way to introduce precise variations in structural and optical properties in a controlled fashion. It has been widely used to modify the band energy as well as structural properties of various semiconductor materials, including GaN[7,8,9,10,11,12]. In this study, we investigate the effect of high-energy 700keV Ni++ ion irradiation on the structural and optical properties of GaN thin films.

**MOCVD growth:**
The GaN thin films in this study were grown on a c-plane sapphire substrate using metal-organic chemical vapor deposition (MOCVD). Trimethyl-gallium (TMGa) and ammonia (NH$_3$) were used as precursors for Ga and N, respectively. The substrate temperature was maintained at 1050°C, and the pressure during the growth process was maintained at 50 Torr. The thickness of the as-grown GaN films was about 5 μm.

**Ion irradiation:**
The GaN layers were then irradiated with 700keV Ni++ ions using a High Energy 5 MV Pelletron Tandem Accelerator. The irradiation was performed with three different doses of $10^{13}$, $10^{14}$, and $10^{15}$ ions/cm$^2$, with a beam current of 50 nA, and the irradiation time was adjusted to achieve the desired fluence.

**X-ray diffraction measurements:**

The X-ray diffraction measurements were performed on as grown as well as irradiated layers using a Panalytical Emperyan diffractometer with Cu Kα radiation (λ = 1.54191Å) in the 2θ range of 30° to 80°. Numerous omega–2Theta rocking curves were recorded and FWHM were extracted. X-ray diffraction spectra of as-grown and 700 keV Ni++ ion irradiated GaN layers are shown in Figure 1. XRD of the as grown layer (Fig 1(a)) shows sharp GaN peaks at 34.55 and 72.91 corresponding to (0002) and (0004) directions, respectively. Our measurements are in agreement with previously reported results[13,14]. Fig. 1 (b), (c) and (d) show the XRD spectra of the irradiated layers. Intensity of the GaN peaks decreases steadily with the increase in ion fluences. FWHM of GaN (0002) peak is found to be increasing from 345.5 arcsec to 1021.2 arcsec.

The broadening of (0002) with irradiation can be attributed introduction of point defects due to displacement of atoms from their lattice sites, creating vacancies, interstitials, and dislocations in the crystal structure[15,16,17]. These defects act as scattering centers for X-rays, leading to a reduction in the coherence length of the crystal i.e. reducing the size of crystalline domains resulting in an increase in the width of the diffraction peak. Additionally, ion irradiation also induces strain in the crystal lattice due to the difference in atomic size and energy between the ions and the host atoms, which contributes to peak broadening. The lattice strains introduced by irradiation cause the peak position to shift. In present case, (0002) peak shifts from 34.55 in as-grown layer to 34.50 in $10^{15}$ fluenced layer. The intensity of GaN (101) peak also increases with increased fluences indicating a relative increase in the concentration of this crystalline phase due to irradiation.

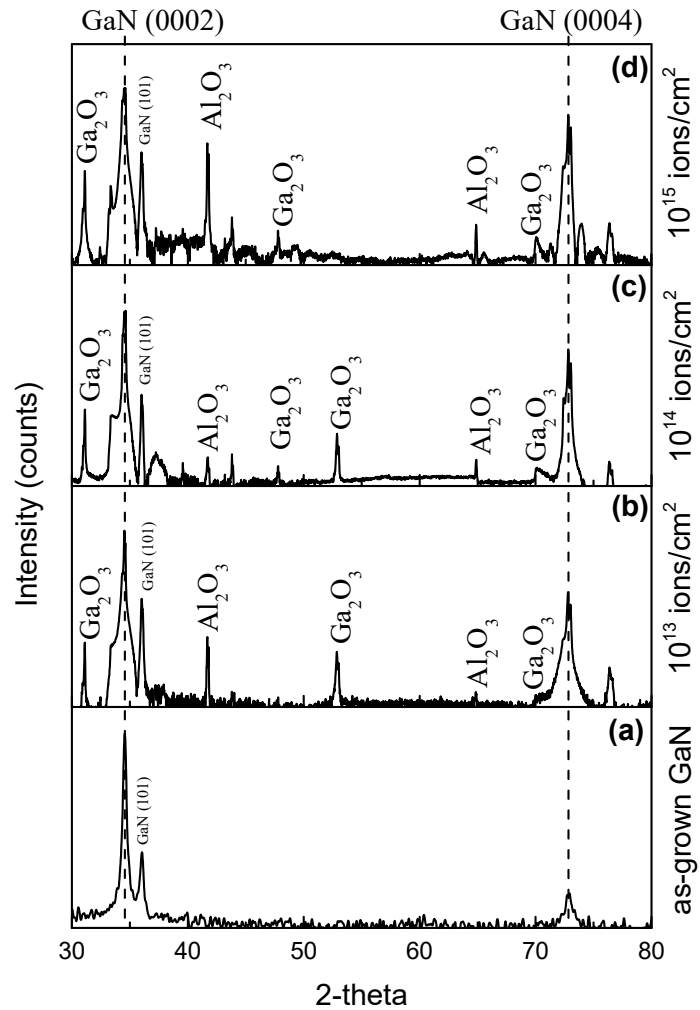

**Figure 1:** X-ray diffraction spectra of (a) as-grown and 700keV Ni+++ ion irradiated GaN layers at a fluence of (b)$10^{13}$, (c) $10^{14}$ and (d) $10^{15}$ ions/cm$^2$, respectively.

The intensity of sapphire substrate peaks increases with increased fluence. $Ga_2O_3$ peaks appearing at 31.11, 52.92, 47.83 and 70.06 can be attributed to interface diffusion at $GaN/Al_2O_3$ interface. Similar results about interface diffusion have been previously reported in literature[18,19,20,21,22].

**UV-visible optical absorption spectroscopy:**

The optical properties of the GaN films were analyzed using a Perkin Elmer Lambda 950 spectrophotometer in the wavelength range of 200-900 nm. The optical absorption coefficient (α) was calculated from the measured transmittance (T) using the following equation:

$$\alpha = -\frac{\ln(T)}{d} \quad (1)$$

Where d is the thickness of the GaN film. The thickness of the as-grown and irradiated films was measured using spectroscopic Ellipsometry measurements. The bandgap energy (Eg) of the GaN films was estimated from the Tauc plot[23,24,25,26,27] which is based on the relationship between the absorption coefficient and the photon energy (hν) near the bandgap:

$$(\alpha h\nu)^2 = A(h\nu - E_g) \quad (2)$$

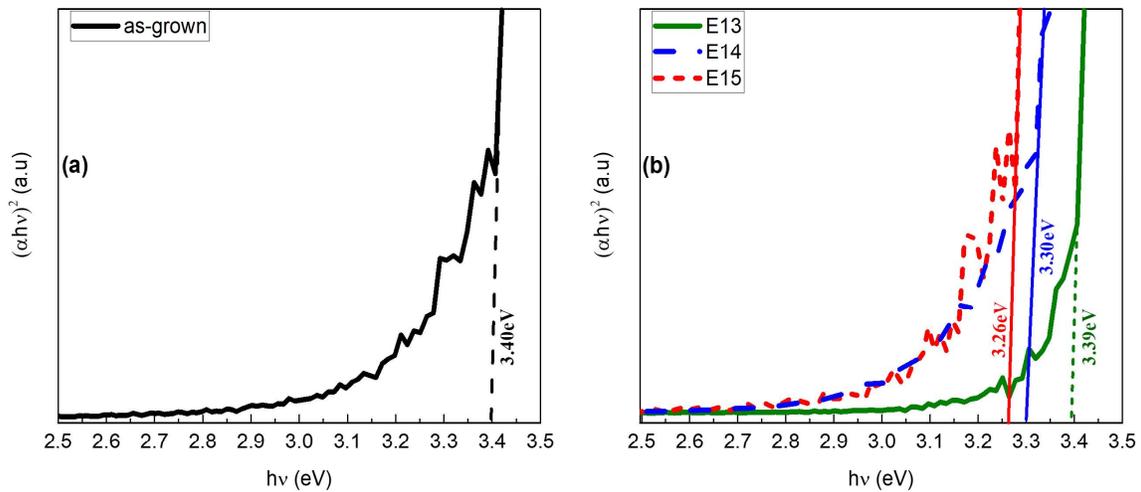

**Figure 2:** Tauc plots for (a) as-grown and (b) 700keV Ni++ ion irradiated GaN layers.

Where A is a constant, and Eg is the bandgap energy.

The measured UV-Vis spectra in energy range of 2.5eV to 3.5eV for the as-grown as well as the ion-irradiated GaN layers are shown in Figure 2. The slopes of the absorption coefficient (ahn)2 vs. photon energy of the as-grown and Ni++ ion irradiated GaN films are also shown in Fig. 2. The absorption edge of the as-grown GaN layer was around 364 nm, corresponding to a bandgap energy of 3.4 eV. After ion irradiation, the absorption edge shifted towards longer wavelengths, indicating a decrease in the bandgap energy of the GaN film. The bandgap energy decreased from 3.4 eV for the pristine GaN film to 3.26 eV for the film irradiated with $10^{15}$ ions/cm$^2$. This decrease in bandgap energy is attributed to the creation of defects and/or the formation of new phases in the GaN film due to ion irradiation, which could introduce new energy states within the bandgap.

Ion irradiation at high energy introduces traps in the deeper levels of the energy bandgap, these defects act as the trapping centers for electrons and holes. This phenomenon results into reduction of the bandgap of the material under study and can be seen as band tailing effects[21], that we also see in Figure 2.

**Conclusions:**
MOCVD grown GaN layers were irradiated with 700 keV Ni++ ion beam at three different irradiation doses of $10^{13}$, $10^{14}$, and $10^{15}$ ions/cm$^2$. X-ray diffraction (XRD) and UV-Vis spectroscopy were used to study the changes in structural and optical properties of irradiated layers. The XRD measurements on irradiated films revealed new peaks indicating the formation of new phases or defects induced by ion irradiation. The intensity of these extra peaks increases with increasing ion dose, suggesting that the density of defects in the GaN film also increases with increasing ion dosage. The UV-Vis measurements indicated a decrease in the bandgap energy irradiated GaN layers from 3.40 eV for the as-grown GaN film to 3.26 eV for the film irradiated with highest dose of $10^{15}$ ions/cm$^2$. The decrease in bandgap can be attributed to the creation of defects and/or the formation of new phases in the GaN film due to the ion irradiation.